\documentclass[epj]{svjour}
\usepackage[dvips]{epsfig}
\usepackage{subfigure}
\usepackage{amsmath,amssymb,bm}
\usepackage[usenames,dvipsnames]{color}
\usepackage{soul}
\setstcolor{Red}
\sethlcolor{Yellow}
\newcommand{\nc}[1]{\newcommand{#1}}
\nc{\its}[1]{\itshape #1 \upshape}
\nc{\mc}[3]{\multicolumn{#1}{#2}{#3}}
\nc{\bc}{\begin{center}}
\nc{\ec}{\end{center}}
\nc{\ig}[1]{\bc \includegraphics{#1} \ec}
\nc{\bo}[1]{\mbox{\boldmath \( #1 \! \! \)  \unboldmath}}
\nc{\beq}{\begin{equation}}
\nc{\eeq}{\end{equation}}
\nc{\bew}{\begin{eqnarray}}
\nc{\eew}{\end{eqnarray}}
\nc{\bs}{\begin{subeqnarray}}
\nc{\es}{\end{subeqnarray}}
\nc{\nnn}{\nonumber \\}
\nc{\Mvariable[1]}{\; #1}
\nc{\f}[2]{\frac{#1}{#2}}
\nc{\td}[2]{\f{d #1}{d #2}}
\nc{\pd}[2]{\f{\partial #1}{\partial #2}}
\nc{\suli}{\sum\limits}
\nc{\proli}{\prod\limits}
\nc{\ili}{\int\limits}
\nc{\sr}[2]{\stackrel{#1}{#2}}
\nc{\dps}{\displaystyle}
\nc{\ket}[1]{\left| #1 \right>}
\nc{\bra}[1]{\left< #1 \right|}
\nc{\bracket}[2]{\left< #1 \right| \left. \! #2 \right>}
\nc{\norm}[1]{\left\| #1 \right\|}
\nc{\lndm}[1]{\pd{^{#1} \ln{\det{M}}}{\mu^{#1}}}
\nc{\pdmm}[1]{M^{-1} \pd{^{#1} M}{\mu^{#1}}}
\nc{\pdm}{M^{-1}\pd{M}{\mu}}
\nc{\trac}[1]{\mbox{Tr}\left(#1\right)}
\nc{\fb}{\color{blue}}
\nc{\fr}{\color{red}}
\nc{\muh}{\hat \mu}
\nc{\nuh}{\hat \nu}
\nc{\rhoh}{\hat \rho}
\nc{\sigmah}{\hat \sigma}
\def \beq{\begin{equation}}
\def \eeq{\end{equation}}
\def \beqa{\begin{eqnarray}}
\def \eeqa{\end{eqnarray}}
\def \lt{\left}
\def \rt{\right}
\def \c{\chi}

\def \bx{\bm x}

\def \npb{{Nucl.\ Phys.\ B}}
\def \npsl{{Nucl.\ Phys.\ Proc.\ Suppl.\ }}

\def \prd{{Phys.\ Rev.\ D}}

\def \prl{{Phys.\ Rev.\ Lett.\ }}

\def \etal{{\sl et al.}}

\begin{document}
\title{Taste symmetry breaking at finite temperature}
\author{Edwin Laermann and Fabrizio Pucci}

\institute{Fakult\"at f\"ur Physik, Universit\"at Bielefeld, D-33615 Bielefeld, Germany}

\abstract{The breaking of the taste symmetry is studied in the temperature range between 140 MeV to 550 MeV.
In order to investigate this violation we have calculated the screening masses of the various taste states
fitting the exponential decay of the spatial correlators. The computation has been performed
using dynamical $N_f$ = 2+1 gauge field configurations generated with the  p4 staggered action along the Line
of Constant Physics (LCP) defined by a pion mass $m_\pi$ of approximately $220$ MeV and the kaon mass $m_K$
equals $500$ MeV.  For temperatures below the transition an agreement with the predictions of the staggered
chiral perturbation theory has been found and no temperature effect can be observed on the taste violation. Above
the transition the taste splitting still shows an $\mathcal{O}(a^2)$ behavior but with a temperature dependent slope. In addition
to the analysis done for the pion multiplet we have performed an analogous computation for the \emph{light-strange} and \emph{strange} mesons and also looked at the scalar, vector and axial vector channels to understand how the multiplets split at finite temperature.
Finally the temperature dependence of the pion decay constant $f_{\pi}$ is investigated to get further information about the chiral symmetry restoration.
\newline
}

\authorrunning{Edwin Laermann and Fabrizio Pucci}
\titlerunning{Taste symmetry breaking at finite temperature}
\date{\bc\today\ec}
\maketitle
\section{Introduction}          \label{se.intro}
One of the most famous 
problems related to the lattice discretization of
quantum chromodynamics (QCD) is the presence of unphysical fermionic modes
called doublers. Several methods have been proposed in the literature to avoid this
proliferation and to reduce their number. One of these proposals is the well known and
widely used staggered fermion formulation introduced for the first time in
\cite{Kogut:1974ag} by Kogut and Susskind. Indeed after the staggering, the
sixteen doublers reduce to four species that are usually called tastes to
distinguish them from the ordinary flavors.
\newline
This construction has become popular in simulations of lattice
QCD, in particular at finite temperature because it is computationally
cheaper and preserves a remnant of the chiral symmetry. However, since
the doubling problem is not completely solved, to reduce further the tastes and
restore the correct number of degrees of freedom a rooting procedure has to be
introduced. The validity of this technique has long been debated in the literature
\cite{Bunk:2004br,Adams:2004mf,Maresca:2004me,Shamir:2004zc,Neuberger:2004be,Bernard:2005gf,Bernard:2004ab,Durr:2004ta,Creutz:2007yg,Creutz:2007rk,Bernard:2006vv,Creutz:2007pr}
and even if a definite conclusion is still missing, numerical evidence seems
to confirm its correctness namely that the non-local terms appearing at finite
lattice spacing become irrelevant in the continuum limit.\newline
At non-zero lattice spacing, due to the fact that for staggered fermions
the spin and the taste degrees of freedom are mixed, a breaking of the continuum taste
symmetry occurs and as a consequence only one of the sixteen pions is a true
Goldstone boson. The other fifteen states split
according to their transformation properties under the symmetry group preserved by the
staggered formulation \cite{Golterman:1985dz,Aubin:2004fs,Bazavov:2010pg}.
Since these taste violations contribute to the cut-off dependence of physical
quantities computed on the lattice, several ways have been introduced and applied
 in the literature to reduce them and to improve the staggered action. In fact, in the p4, aqstad,
stout and HISQ actions this improvement was obtained by using some variants of the gauge-link smearing
\cite{Heller:1999xz,Karsch:2000ps,Orginos:1999cr,Aoki:2005vt,Follana:2006rc}.
\newline
In order to have a better comprehension of the impact of these cut-off effects on
physical observables it is important to understand how
taste symmetry is violated at finite lattice spacing. The concrete way to do that
is usually to look at the so called taste splitting 
defined by the difference between
the square of the 
mass of a non-Goldstone pion and the Goldstone one
$\Delta_{\xi} = m_{[\xi]}^2 - m_{\xi_5}^2$\footnote{In the spinor-taste basis we will
indicate with $\gamma$ and $\xi$ the spin and the taste gamma matrices respectively.}. At zero temperature the situation
is quite clear. All the results, see e.g.
\cite{Aubin:2004fs,Bazavov:2010pg,Bazavov:2010ru} for the most recent investigations,
confirm that in the pion sector
the taste splitting goes like $a^2$ where $a$ is the lattice
spacing. Remarkably this behavior can be well described with a tree-level
computation in staggered chiral perturbation theory (S$\chi$PT) \cite{LeeSharpe,AubinBernard,Bernard:2007qf}.
Indeed in the staggered chiral Lagrangian 
taste-breaking four-fermion operators appear,
physically due to the fact that the exchange of gluons with momenta $q\sim \pi/a$
can change the taste of the staggered quarks.
\newline
At finite temperature the situation is not so well understood. S$\chi$PT is
formulated at zero temperature and can not help in the understanding of taste symmetry
breaking because in principle the structure
of the chiral Lagrangian could be modified at finite $T < T_c$,
where $T_c$ is the temperature of approximate restoration of chiral symmetry,
possibly leading to
temperature effects. For temperatures above the transition we expect the breakdown of S$\chi$PT.
\newline
The aim of this investigation is thus to go further into the problem and
to study the dependence of taste breaking on lattice spacing
as well as temperature. In order to extract this information we will calculate
the screening masses of the different taste components of the pion multiplet at finite
temperature in $2+1$ flavor QCD using an improved p4 \cite{Heller:1999xz} staggered
fermion action
at bare quark masses which correspond to a 
Goldstone pion mass of $220$ MeV and
a kaon mass of $500$ MeV at zero temperature.
The temperature will range from about $140$ MeV
to $550$ MeV and our study will be performed at two different sets of the
lattice spacing corresponding to $N_\tau= 6$ and $8$. Our analysis will not be restricted
to the pion sector, we have studied also the scalar, the axial-vector and the vector channels
to obtain useful indications about taste violations also in those channels. Additionally we
address the effect of the flavors on the taste breaking, analyzing some local and
non-local
pseudoscalar 
operators in light-strange ($\bar{u}s$) and strange-strange ($\bar{s}s$) channels.
\newline
The screening mass analysis is relevant not only for the understanding of the taste
symmetry breaking  but also for the study of the chiral symmetry. It is a well known fact
that an advantage of the staggered formulation is to preserve a $U(1)_{\epsilon}$ subgroup\footnote{This subgroup should not be confused with the
anomalous $U(1)_A$ of continuum QCD.}
of the chiral symmetry in the continuum. Since this subgroup relates different classes of mesonic operators,
in the symmetric phase it predicts the degeneracy between certain channels.
Above the transition its restoration demands that at zero quark mass
 \begin{equation}C(z)_{AV}=(-1)^z C(z)_{V}\label{itsrestoration}\end{equation}
\noindent
where $C_{V}$ and $C_{AV}$ are the correlators in the vector ($\gamma_{\mu}$) and axial vector ($\gamma_5 \gamma_{\mu}$)
channels respectively.
\newline
Finally there is a general motivation to study screening masses since from them one can
extract information about some non-perturbative properties of the quark - gluon
plasma. Indeed the inverse of the screening mass describes the distance beyond which the
effect of a test hadron put into the medium is screened.
\newline
The plan of the paper is the following: after an introduction to staggered mesons
(section~\ref{se.free}) where we briefly discuss the symmetry properties of the staggered
formulation both at zero and finite temperature, we present a summary of the predictions
of staggered chiral perturbation theory on the taste breaking (section~\ref{se.SCHPT}).
Section (\ref{se.light}) is devoted to the presentation of the results regarding the
screening masses of the different mesonic states at finite temperature before in (\ref{se.summary}) we will summarize our results, discuss some open problems
and suggest possible directions for a future investigation.

\section{Staggered Mesons}                             \label{se.free}
The symmetry properties of the system are fundamental to the analysis of the
correlation functions and the determination of the screening masses. In fact all
the states can be classified according to their transformation properties under
the subgroup preserved by the lattice discretized theory. For these reasons in this section we review
some basic and known facts about the classification of the staggered meson correlators
starting with zero temperature \cite{Golterman:1985dz,Golterman:1984dn} and then
passing to the finite temperature case \cite{Gupta:1999hp}.
\newline
In the continuum limit the symmetry group of one staggered fermion is $SU(2) \times U(4)$
with the first factor associated to the rotational symmetry and the second one with the taste symmetry.
At non zero lattice spacing this invariance group is broken down to the isometries group of the lattice.
Since we are looking at zero momentum states with a fixed quark number, staggered mesons can be classified
according to their transformation properties under a subgroup of the RF group\footnote{The rest frame group
is the invariance group of the transfer matrix}, the so called geometric time slice group (GTS) that is generated by the
spatial inversion $I_s$, the taste transformations $\Xi_{\mu}$ and the rotations of $\pi/2$ angle $R^{kl}$.
The staggered quark field $\chi($\textbf{x}$)$ defined at the point $\textbf{x}$ = $(x,y,z,t)$ on a lattice
with spacing $a$ forms an
8 dimensional irreducible representation (8-irrep) of the GTS. Under the subgroup of the discretized version of the cubic rotation group
$O_h$ it decomposes as $A_1^+ \oplus A_1^- \oplus F_1^+ \oplus F_1^-$ where $A_1^{\pm}$ and $F_1^{\pm}$ are, respectively,
the trivial and the three dimensional vector representation with parity $\pm$. To construct the meson states
one has to take the tensor product of two staggered fields and just by counting one obtains sixty-four states
that, following \cite{Golterman:1985dz}, we list in Table \ref{tab.Operator}. At a first sight
the meson operators can be divided in four classes: the local operators ($n=1..4$ in Table \ref{tab.Operator})
written as
\begin{equation}\mathcal{M}^{0-L} = \phi(\textbf{x}) \bar{\chi}(\textbf{x})\chi(\textbf{x}),\end{equation}
 with $\phi(\textbf{x}$) being a phase factor depending on the choice of the channel,
where the two staggered fields sit at the same lattice point; the one-link operators ($n=5..10$) written schematically as
\begin{equation}\mathcal{M}^{1-L} = \phi(\textbf{x}) \bar{\chi}(\textbf{x})\Delta_i \chi(\textbf{x})\end{equation} where one has to introduce the shift operator as
\begin{equation}\Delta_i\, \chi(\textbf{x}) = 1/2 ( \chi(\textbf{x}+ \hat{i}) +  \chi(\textbf{x}- \hat{i}))\end{equation}
and put the quark and the anti-quark in two next-neighbor lattice points;
and finally the two and three link
operators ($n=11..16$ and $n=17..20$ respectively) defined as
 \begin{equation}\mathcal{M}^{2-L} =\epsilon_{ijk} \phi(\textbf{x}) \bar{\chi}(\textbf{x})\Delta_i \Delta_j\chi(\textbf{x})\end{equation}
 \begin{equation}\mathcal{M}^{3-L} = \phi(\textbf{x}) \bar{\chi}(\textbf{x})\Delta_1 \Delta_2 \Delta_3 \chi(\textbf{x})\end{equation}
  where quark and anti-quark are separated by two and three links. More specifically, all the states can be classified according to the value  $\textbf{\textbf{\emph{r}}}^{\sigma_s, \sigma_{123}}$ of the representation of GTS in which they lie, where $\textbf{\emph{r}}$ is the dimension of the irrep., and $\sigma_s$ and $\sigma_{123}$ are, respectively,
the eigenvalues of $I_s$, the spatial inversion, and of $X_{123}$, the
parity under spatial taste inversion. Just for completeness, in
the fourth column of
Table \ref{tab.Operator} we have also indicated
the meson description in the original language of Kogut and Susskind: using the spin $\otimes$ taste basis the
staggered mesons can be written as  $M = \bar\psi\lt(\Gamma^D\otimes\Gamma^T\rt)\psi$ where $\psi$ has four Dirac and four taste
components and is defined on a coarse lattice with spacing $2 a$. From the
fifth column one can read off the $J_R^{PC}$ quantum numbers of the
lowest corresponding continuum states at zero temperature
and finally in the last column the particle identification of those states is given.

\begin{table*}
\begin{center}
\begin{tabular}{cccccc}
\hline
  & & & & & \\
N & Operator & $r^{\sigma_s \sigma_{123}}$ & $\Gamma^D \times \Gamma^F$ & State & Particle \\

\hline
  & & & & & \\
1 & $\bar{\chi} \chi$                    & $1^{++}$ & $1 \otimes 1$ &   $0_S^{++}$       &  $f_0$         \\
  &                                      &          & $\gamma_4 \gamma_5 \otimes \xi_4 \xi_5$ &  $0_A^{-+}$        & $\pi$          \\
2 & $\eta_{4} \zeta_{4} \bar{\chi} \chi$ & $1^{+-}$ & $\gamma_4 \otimes \xi_4 $ & $0_A^{+-}$         &  -        \\
  &                                      &          & $ \gamma_5 \otimes \xi_5$ & $0_A^{-+}$         &   $\pi$       \\
3 & $\eta_{i} \epsilon \zeta_{i} \bar{\chi} \chi$ & $3^{+-}$ & $ \gamma_i \gamma_5 \otimes \xi_i \xi_5$ & $1_A^{++}$ & $a_1$ \\
 &                                      &          & $ \gamma_i \gamma_4 \otimes \xi_i \xi_4$ & $1_A^{--}$  &    $\rho$      \\
4 & $\eta_{4} \zeta_{4} \eta_{i} \epsilon \zeta_{i} \bar{\chi} \chi$ & $3^{++}$ & $ \gamma_j \gamma_k \otimes \xi_j \xi_k$ & $1_A^{+-}$ & $b_1$ \\
 &                                      &          & $ \gamma_i \otimes \xi_i $ & $1_A^{--}$         &  $\rho$        \\
5 & $\bar{\chi} \eta_i \Delta_i \chi$  & $3^{-+}$ & $ \gamma_i \otimes 1$  & $1_S^{--}$ & $\omega$ \\
 &                                      &          & $ \gamma_j \gamma_k \otimes \xi_4 \xi_5$ &  $1_A^{+-}$        &  $b_1$        \\
6 & $\eta_{4} \zeta_{4} \bar{\chi} \eta_i \Delta_i \chi$ & $3^{--}$ & $ \gamma_i \gamma_4 \otimes \xi_4$  & $1_A^{--}$ & $\rho$ \\
 &                                      &          & $ \gamma_i \gamma_5 \otimes \xi_5$ & $1_A^{++}$         &  $a_1$        \\
7 & $\bar{\chi} \epsilon \zeta_{i} \Delta_i \chi$ & $3^{--}$ & $ \gamma_5 \otimes \xi_i \xi_5$  & $0_A^{-+}$ & $\pi$ \\
 &                                      &          & $ \gamma_4 \otimes \xi_i \xi_4$ & $0_A^{+-}$         &    -      \\
8 & $\eta_{4} \zeta_{4} \bar{\chi} \epsilon \zeta_{i} \Delta_i \chi$ & $3^{-+}$ & $ \gamma_4 \gamma_5 \otimes \xi_j \xi_k$  & $0_A^{-+}$ & $\pi$ \\
 &                                      &          & $ 1 \otimes \xi_i $ &  $0_A^{++}$        &  $a_0$        \\
9 & $\eta_{i} \epsilon \zeta_{i} \bar{\chi} \zeta_{j} \Delta_j \chi$ & $6^{--}$ & $ \gamma_k \gamma_4 \otimes \xi_i \xi_5$  & $1_A^{--}$ & $\rho$ \\
 &                                      &          & $ \gamma_k \gamma_5 \otimes \xi_i \xi_4$ & $1_A^{++}$         &     $a_1$     \\
10 & $\eta_{4} \zeta_{4} \eta_{i} \epsilon \zeta_{i} \bar{\chi} \zeta_{j} \Delta_j \chi$ & $6^{-+}$ & $ \gamma_k \otimes \xi_j \xi_k$  & $1_A^{--}$ & $\rho$ \\
 &                                      &          & $ \gamma_i \gamma_j \otimes \xi_i$ & $1_A^{+-}$         &     $b_1$     \\
11 & $\epsilon_{ijk} \bar{\chi} \eta_i \Delta_i ( \eta_j \Delta_j \chi ) $ & $3^{++}$ & $ \gamma_i \gamma_j \otimes 1$  & $1_S^{+-}$ & $h_1$ \\
 &                                      &          & $ \gamma_k \otimes \xi_4 \xi_5$ & $1_A^{--}$         &  $\rho$        \\
12 & $\epsilon_{ijk} \eta_{4} \zeta_{4} \bar{\chi} \eta_i \Delta_i ( \eta_j \Delta_j \chi ) $ & $3^{+-}$ & $ \gamma_k \gamma_5 \otimes  \xi_4$  & $1_A^{++}$ & $a_1$ \\
 &                                      &          & $ \gamma_k \gamma_4 \otimes \xi_5$ & $1_A^{--}$         &   $\rho$       \\
13 & $\epsilon_{ijk} \bar{\chi} \zeta_i \Delta_i ( \zeta_j \Delta_j \chi ) $ & $3^{++}$ & $ 1 \otimes \xi_i \xi_j$  & $0_A^{++}$ & $a_0$\\
 &                                      &          & $ \gamma_4 \gamma_5 \otimes \xi_k$ & $0_A^{-+}$         &   $\pi$       \\
14 & $\epsilon_{ijk} \eta_{4} \zeta_{4} \bar{\chi} \zeta_i \Delta_i ( \zeta_j \Delta_j \chi ) $ & $3^{+-}$ & $ \gamma_4 \otimes \xi_k \xi_5$  & $0_A^{+-}$ & - \\
 &                                      &          & $ \gamma_5 \otimes \xi_k \xi_4$ & $0_A^{-+}$         &  $\pi$        \\
15 & $\eta_k \zeta_k \bar{\chi} \eta_i \Delta_i ( \zeta_j \Delta_j \chi ) $ & $6^{++}$ & $ \gamma_i \gamma_k \otimes \xi_j \xi_k$  & $1_A^{+-}$ & $b_1$ \\
 &                                      &          & $ \gamma_j \otimes \xi_i $ & $1_A^{--}$         &   $\rho$       \\
16 & $\eta_{4} \zeta_{4} \eta_k \zeta_k \bar{\chi} \eta_i \Delta_i ( \zeta_j \Delta_j \chi ) $ & $6^{+-}$ & $ \gamma_j \gamma_5 \otimes \xi_i \xi_5$ & $1_A^{++}$ & $a_1$ \\
 &                                      &          & $ \gamma_j \gamma_4 \otimes \xi_i \xi_4$ & $1_A^{--}$         &  $\rho$        \\
17 & $\bar{\chi} \eta_1 \Delta_1 ( \eta_2 \Delta_2 ( \eta_3 \Delta_3 \chi ))$ & $1^{-+}$ & $ \gamma_4 \gamma_5 \otimes 1$ & $0_S^{-+}$ & $\eta^{'}$ \\
 &                                      &          & $ 1 \otimes \xi_4 \xi_5$ & $0_A^{++}$         &  $a_0$        \\
18 & $\eta_{4} \zeta_{4} \bar{\chi} \eta_1 \Delta_1 ( \eta_2 \Delta_2 ( \eta_3 \Delta_3 \chi ))$ & $1^{--}$ & $ \gamma_5 \otimes \xi_4$ & $0_A^{-+}$ & $\pi$ \\
 &                                      &          & $ \gamma_4 \otimes \gamma_5$ & $0_A^{+-}$         &    -      \\
19 & $\eta_i \epsilon \zeta_i \bar{\chi} \eta_1 \Delta_1 ( \eta_2 \Delta_2 ( \eta_3 \Delta_3 \chi ))$ & $3^{--}$ & $ \gamma_i \gamma_4 \otimes \xi_i \xi_5$ & $1_A^{--}$ & $\rho$ \\
 &                                      &          & $ \gamma_i \gamma_5 \otimes \xi_i \xi_4$ & $1_A^{++}$         &   $a_1$       \\
20 & $\eta_{4} \zeta_{4} \eta_i \epsilon \zeta_i \bar{\chi} \eta_1 \Delta_1 ( \eta_2 \Delta_2 ( \eta_3 \Delta_3 \chi ))$ & $3^{-+}$ & $ \gamma_i \otimes \xi_j \xi_k$ & $1_A^{--}$ & $\rho$ \\
 &                                      &          & $ \gamma_j \gamma_k \otimes \xi_i$ & $1_A^{+-}$         &  $b_1$        \\
& & & & & \\\hline
\end{tabular}
\end{center}
\caption{Irreducible representations and their operators for mesonic states. Here $\Delta_i f(x,t) = \frac{1}{2}\, [ f(x+\hat{i},t) + f(x-\hat{i},t)
]$ while $\eta_{\mu}(x) = (-1)^{x_1 + ... x_{\mu-1}}$, $\zeta_{\mu}(x) = (-1)^{x_{\mu+1} + ... x_{4}}$ and $\epsilon(x) = (-1)^{x_1 +  ... x_{4}}$.
}
\label{tab.Operator}
\end{table*}

\noindent
At non-zero temperature the situation is different due to the fact that the temporal direction is special since it is related to
the temperature through the relation $T =  1/a N_{\tau}$
where $N_\tau$ is the number of lattice points in that direction.
For spatial correlation functions, which we are going to analyze, the subgroup of $O_h$
preserved in the zero temperature case is now broken since certain $\pi/2$ rotations (for example in the $x-t$
and $y-t$ plane) are no more symmetries of the lattice discretization. As a consequence, the mesonic states have to be
classified differently from Table \ref{tab.Operator}.  The new invariance subgroup is $D_4^h$,
the discretized
version of the cylinder symmetries with 16 elements divided into eight one-dimensional
representations labeled by $A_{1}^{\pm}, A_{2}^{\pm}, B_{1}^{\pm}, B_{2}^{\pm}$ and two
two-dimensional irreps $E^{\pm}$. Following \cite{Gupta:1999hp} it is easy to understand
how the irreps of $O_h$ reduce under $D_4^h$. Basically the one dimensional irrep of $O_h$ continues to be a 1-irrep of $D_4^h$ while
 the vector of $O_h$ decomposes into a singlet plus the two dimensional representation $E$. One can now do the tensor product
 of two staggered quark fields and find in which way the mesons transform under the invariance group at finite $T$. This has already been done
 by S. Gupta in \cite{Gupta:1999hp} and summarizing his results we have that the three dimensional vector irreps, which in Table \ref{tab.Operator} are labeled
 by the numbers $n=3..9,11..14$, and $19,20$, split into a two plus a one dimensional
representation. Concerning the tensor irreps. of dimension $6$, labeled by $n=9,10,15,16$, they split in four plus two dimensional representations.
\newline
If the temperature is further increased and approaches the infinite temperature limit,
compared to the finite temperature case, more degeneracies will develop
as in this limit the theory approaches the free case.
Indeed to the extent that dimensional reduction is
a good approximation, one sees an approximate $C_v^4$ symmetry\footnote{The symmetry group of the dimensionally reduced theory $C_v^4 \simeq D_4$
is the group of the isometries of a square.},
 thus the representations $A_1^+, E^+, A_1^-$ become degenerate with
$A_2^-, E^-, A_2^+$ respectively and as a consequence the correlators fall into only very few multiplets.
\newline
In this work we study the local and non-local mesons built from $\bar ud$, $\bar us$ and $\bar ss$
flavor combinations.
In flavor singlet channels like e.g. $\bar s s$ only the quark-line connected part has been
computed and the computationally demanding disconnected part has been neglected.
 The connected part of the staggered meson screening correlators,
projected to zero transverse momentum, ${\bm p}_\perp = (p_x,p_y)$, and to zero
(boson) Matsubara frequency $\omega_n$, are obtained as
\small
\beq
C(z) = \sum_{x,y,\tau}\, \, \sum_{\textbf{v},\textbf{w} \in \mathfrak{L}} \phi(\bx)
\lt\langle\lt( M^{-1}_{{\bx}+\textbf{v},\bm{0}} U_\textbf{v}(\textbf{x}) \rt) \lt(M^{-1}_{{\bx}\textbf{w}} U_\textbf{w}(\textbf{-w}) \rt)^\dagger  \rt\rangle,
\eeq
\normalsize
where $M^{-1}_{{\bm 0}\bx}$ is the full staggered quark propagator from ${\bm 0}$ to $\bx$, $\phi(\textbf{x})$ is a phase depending
on the choice of the channel and $\mathfrak{L}$ = $\{0\}$ for the local correlators, $\mathfrak{L}$ = $\{ \pm \hat{i} \}$ for the one-link correlators and is equal to  $\{\pm \hat{i} \pm \hat{j}\}$ and $\{ \pm\, \hat{x} \pm \hat{y} \pm \hat{\tau}\}$  respectively for the two-link and the three link correlation functions\footnote{$\hat{i}$ and $\hat{j}$ $\in$  $\{\hat{x},\hat{y},\hat{\tau}\}$ }. Thus while for the computation of the screening mass of the local meson only one inversion
of the fermion matrix is needed, for the one-link, two-link and three-link operators we need three, five and nine inversions respectively that make the computation more demanding
compared
to the local operators. Since a staggered fermion meson correlator, in general, contains two different mesons
with opposite parity \cite{Golterman:1985dz} one has to parametrize it as
\bew
C(z) = && A_{NO} \cosh\lt[ M_- \lt( z - \frac{N_s}{2} \rt)\rt] \nnn
& - & (-1)^z A_O \cosh\lt[ M_+ \lt( z - \frac{N_s}{2} \rt)\rt] .
\label{eq.cor-par}
\eew
According to our phase convention, for the local and two-link operators $M_-$ ($M_+$) corresponds to the screening mass of the lightest
negative (positive) parity state and comes as the non-oscillating, NO,
(oscillating, O,) part of the screening correlator and
viceversa for the one and three-link operators\footnote{Note that both amplitudes are positive \cite{altmeyer}, $A_{NO}, A_O \ge 0$}.

\section{Staggered Chiral Perturbation Theory}                             \label{se.SCHPT}
In this section we review some elements of staggered chiral perturbation theory ( S$\chi$PT) and more in detail its prediction regarding taste symmetry violation. The chiral lagrangian of the S$\chi$PT has been introduced in \cite{LeeSharpe} for a single staggered flavor and generalized to multiple flavors in \cite{AubinBernard,Bernard}. It describes the low energy dynamics of pseudo-Goldstone bosons near the continuum and chiral limit and provides a systematic method to discuss the consequences of chiral symmetry breaking in the framework of the staggered formulation of lattice QCD.
\newline
Starting with an $SU(3)$ gauge theory with $n$ flavors of staggered fermions,
in the combined chiral-continuum limit
 the theory possesses an $SU(4n)_L  \times SU(4n)_R$ chiral symmetry which is
spontaneously broken to $SU(4n)_V$. The 16$n^2$-1 mesonic fields that arise from the breaking can be collected into a unitary matrix

\begin{equation}\Sigma = \texttt{exp} \left( i \Phi / f_{\pi} \right)\end{equation}
\noindent
with
\begin{center}
$\Phi$= $
\left(\begin{array}{cccc}
  U & \pi^+ & K^+ & \cdots \\
  \pi^- & D & K^0 & \cdots \\
  K^- & \bar{K}^0 & S & \cdots \\
  \vdots & \vdots & \vdots & \ddots
\end{array}\right)
$
\end{center}
\noindent
where $U = \sum_{a=1}^{16} U_a T_a$, $\pi^+ = \sum_{a=1}^{16} \pi^+_a T_a$, etc. are  $4\times4$ submatrices expressed in the Hermitian basis\footnote{$\xi_{\mu}$ are the Dirac gamma matrices and $\xi_I= \mathbb{I}_{4\times4}$}

\begin{equation}T_a = { \xi_5, i \xi_{\mu 5 }, i \xi_{\mu \nu}, \xi_{\mu}, \xi_I }.\end{equation}
\noindent
and $f_{\pi}$ is the pion decay constant. Using the usual power-counting scheme to derive the staggered chiral lagrangian

\begin{equation}p^2/\Lambda^{2}_{QCD} \approx m/\Lambda_{QCD} \approx a^2 \Lambda^{2}_{QCD}\end{equation}
\noindent
one obtains to lowest order

\begin{displaymath} {\cal L}_{S\chi PT} = \frac{f_{\pi}^2}{8} Tr ( \partial_{\mu} \Sigma \partial_{\mu} \Sigma^{\dagger} ) - \frac{\mu f_{\pi}^2}{4} Tr ( M^{\dagger} \Sigma + M \Sigma^{\dagger} ) + \end{displaymath}\begin{equation} \frac{2}{3} m_0^2 \left( U_I^2 + D_I^2 +S_I^2 + \cdots \right) + a^2 V ( \Sigma )\label{chila}\end{equation}
\noindent
where

\begin{center}
$M$= $
\left(\begin{array}{cccc}
  m_u\, \mathbb{I} & 0 & 0 & \, \, \cdots \\
  0 & m_d\, \mathbb{I} & 0 & \, \, \cdots \\
  0 & 0 & m_s\, \mathbb{I} & \, \, \cdots \\
  \vdots & \vdots & \vdots & \ddots
\end{array}\right)
$
\end{center}
\noindent
is the quark mass matrix,
$m_0$ models the effect of the anomaly
 and $V(\Sigma)$ is the taste-breaking potential. This potential arises from the four-fermion
operators in the quark effective action and is a linear combination of operators

\begin{equation}- V (\Sigma) = \sum_i C_i O_i\end{equation}
\noindent
with $C_i$ six unknown low energy effective constants (LECs). More in detail the operators that enter in the potential read as

\small
\begin{equation}
O_1 = \texttt{Tr} \left( \Xi_5 \Sigma \Xi_5 \Sigma^{\dagger} \right)
\end{equation}

\begin{equation}O_{2V} = \frac{1}{4} \sum_{\mu} \left[ \texttt{Tr} \left( \Xi_{\mu} \Sigma \right) \texttt{Tr} \left( \Xi_{ \mu} \Sigma \right) + h.c. \right]\end{equation}

\begin{equation}O_{2A} = \frac{1}{4} \sum_{\mu} \left[ \texttt{Tr} \left( \Xi_{\mu 5} \Sigma \right) \texttt{Tr} \left( \Xi_{\mu 5} \Sigma \right) + h.c. \right]\end{equation}

\begin{equation}O_{3} = \frac{1}{2} \sum_{\mu}\left[ \texttt{Tr} \left( \Xi_{\mu} \Sigma \Xi_{\mu} \Sigma \right) + h.c. \right]\end{equation}

\begin{equation}O_{4} = \frac{1}{2} \sum _{\mu} \left[ \texttt{Tr} \left( \Xi_{\mu 5} \Sigma \Xi_{\mu 5} \Sigma \right) + h.c. \right]\end{equation}

\begin{equation}O_{5V} = \frac{1}{2} \sum_{\mu} \left[ \texttt{Tr} \left( \Xi_{\mu} \Sigma \right) \texttt{Tr} \left( \Xi_{ \mu} \Sigma^{\dagger} \right) + h.c. \right]\end{equation}

\begin{equation}O_{5A} = \frac{1}{2} \sum_{\mu} \left[ \texttt{Tr} \left( \Xi_{\mu 5} \Sigma \right) \texttt{Tr} \left( \Xi_{\mu 5} \Sigma^{\dagger} \right) + h.c. \right]\end{equation}

\begin{equation}O_{6} = \frac{1}{2} \sum_{\mu < \nu } Tr \left( \Xi_{\mu \nu}\, \Sigma\, \Xi_{\mu \nu}\, \Sigma^{\dagger} \right)\end{equation}

\normalsize
\noindent
Here the $4n \times 4n$ matrices $\Xi_T$ are just the generalizations of the $4 \times 4$ taste matrices $\xi_T$. If we focus only on the non-diagonal flavor mesons\footnote{Simulations in which disconnected contributions are not taken into account describe only non-diagonal flavor states $\pi^{\pm}, K^{\pm}..$} it is sufficient to expand the lagrangian to quadratic order in the mesonic fields and
consider only the single-trace terms.
>From the expansion one can read off the tree-level masses of the mesons :
\noindent
\begin{equation}
m_{M_B}^2 = \mu ( m_a + m_b ) + a^2 \Delta_{\xi_B}
\label{eq.splittings}
\end{equation}
\noindent
where the meson $M$ is composed of two quarks $a$ and $b$ and where

\begin{equation}\Delta(\xi_5) = \Delta_{PS} = 0\label{a1}\end{equation}

\begin{equation}\Delta(\xi_{\mu 5}) = \Delta_{A} = \frac{16}{f_{\pi}^2} \left( C_1 + 3 C_3 + C_4 + 3 C_6 \right)\label{a2}\end{equation}

\begin{equation}\Delta(\xi_{\mu \nu}) = \Delta_{T} = \frac{16}{f_{\pi}^2} \left( 2 C_3 + 2 C_4 + 4 C_6 \right)\label{a3}\end{equation}

\begin{equation}\Delta(\xi_{\mu}) = \Delta_{V} = \frac{16}{f_{\pi}^2} \left( C_1 + C_3 + 3 C_4 + 3 C_6 \right)\label{a4}\end{equation}

\begin{equation}\Delta(\xi_I) = \Delta_{I} = \frac{16}{f_{\pi}^2} \left( 4 C_3 + 4 C_4 \right)\label{a5}\end{equation}
\noindent
For the flavor neutral mesons ($U$, $D$, $\cdots$) the situation is more complicated since in the expansion of the two-trace operators some two-point vertices mixing the taste-vector, taste-axial and taste-singlet flavor-neutral states appear and thus the Lagrangian receives a contribution of the form
\begin{eqnarray}
\nonumber  \mathcal{L} = -\frac{16 a^2}{f_{\pi}^2} \delta_V ( U_{\xi_{\mu}} + D_{\xi_{\mu}} + \cdots)^2 +  \\
\nonumber - \frac{16 a^2}{f_{\pi}^2} \delta_{AV} ( U_{\xi_{\mu\, 5}} + D_{\xi_{\mu\, 5}} + \cdots )^2\\
 - \frac{16 a^2}{f_{\pi}^2} \delta_I ( U_{I} + D_{I} + \cdots )^2.
\end{eqnarray}
\noindent
For more details about the taste splitting in such channels we refer to \cite{AubinBernard}.
\newline
One interesting thing to note is that the symmetry group of the lattice theory at $\mathcal{O}(a^2)$ is enlarged in the meson sector
since the potential $\mathcal{V}(\Sigma)$ is rotationally and $SO(4)$ taste invariant. The consequence of this enhancement of
symmetry is that the sixteen taste states instead of falling into eight different irreps of the $GTS$ as predicted from group theory split into only five degenerate multiplets according to the $SO(4)$ symmetry (eq. \ref{a1}-\ref{a5}).
The vanishing of $\Delta(\xi_5)$ is due to the taste non-singlet $U_{\epsilon}(1)$ symmetry which is unbroken by the lattice regulator,
 making the $\gamma_5 \times \xi_5$ meson a true Goldstone boson. This prediction of the splitting has been well confirmed in a series of lattice simulations \cite{Aubin:2004fs,Bazavov:2010pg,Bazavov:2010ru,Bernard:2001av}.\newline
This analysis is strictly valid at temperature zero. Since our aim is to study the taste symmetry violations at finite $T$ we need to understand how the predictions of S$\chi$PT get modified. In
 particular two scenarios seem possible. The first one is that the formal structure of the chiral Lagrangian remains the same as in (\ref{chila}) but the coefficients $C_i$ and $f_\pi$ acquire a temperature dependence. At least this hypothesis could work until
  $T_c$ where the chiral symmetry is restored and chiral perturbation theory breaks down. Indeed, as we will see, in the vicinity of the transition the value of the pion decay constant that "measures" the strength of the chiral symmetry breaking goes to zero rapidly. A second scenario can be described by a different lagrangian since in principle operators
different from the $T=0$ case are admitted in the Symanzik effective theory. This is due to the fact that, as explained in the previous section, the invariance group of the zero temperature theory is broken down to a subgroup.

\section{Results}                          \label{se.light}

\begin{table}[!t]
\begin{center}
\begin{tabular}{|ll|lr|lr|}
\hline
        &       &                   \multicolumn{2}{|c|}{$24^3 \times 6$} &
                  \multicolumn{2}{|c|}{$32^3 \times 8$} \\
$\beta$ & $m_l$ & $T$ & conf. &  $T$ & conf. \\
\hline
3.351 & 0.00591 &  145 & 695 &     &    \\
3.410 & 0.00412 &  175 & 786 &     &    \\
3.430 & 0.00370 &  186 & 875 & 139 & 793\\
3.445 & 0.00344 &  197 & 946 &     &    \\
3.455 & 0.00329 &  200 & 405 &     &    \\
3.460 & 0.00313 &  203 & 729 &     &    \\
3.490 & 0.00290 &  226 & 775 &     &    \\
3.500 & 0.00253 &      &     & 175 & 614\\
3.510 & 0.00259 &  240 & 577 &     &    \\
3.530 & 0.00253 &      &     & 192 & 590\\
3.540 & 0.00240 &  259 & 591 &     &    \\
3.570 & 0.00212 &  281 & 525 & 211 & 738\\
3.585 & 0.00192 &      &     & 219 & 360\\
3.630 & 0.00170 &  326 & 389 &     &    \\
3.690 & 0.00150 &  365 & 448 &     &    \\
3.760 & 0.00130 &  424 & 465 & 318 & 588\\
3.820 & 0.00125 &      &     & 361 & 488\\
3.920 & 0.00110 &      &     & 410 & 481\\
3.920 & 0.00092 &  532 & 480 &     &    \\
4.000 & 0.00092 &      &     & 475 & 493  \\
4.080 & 0.00081 &      &     & 549 & 392  \\

\hline
\end{tabular}
\end{center}
\caption{Coupling constants $\beta$ and light quark masses $m_l$, temperatures $T$ in
MeV and the number of configurations of the given sizes on which screening masses
were computed. The strange quarks have always been choice as $m_s = 10 m_l$.}
\label{tb.runpara}
\end{table}

\begin{figure*}[!ht]
\begin{center}
\subfigure[]{\label{beta343}\includegraphics[scale=0.9]{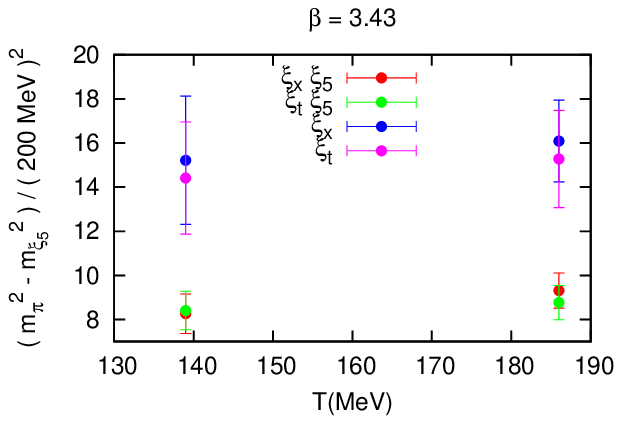}}
\subfigure[]{\label{beta350}\includegraphics[scale=0.9]{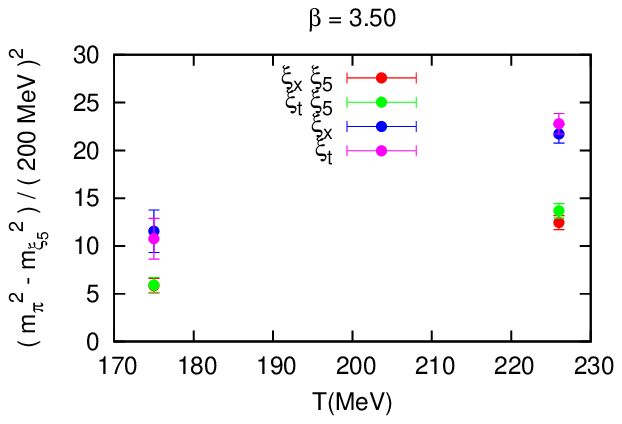}}
\subfigure[]{\label{beta357}\includegraphics[scale=0.9]{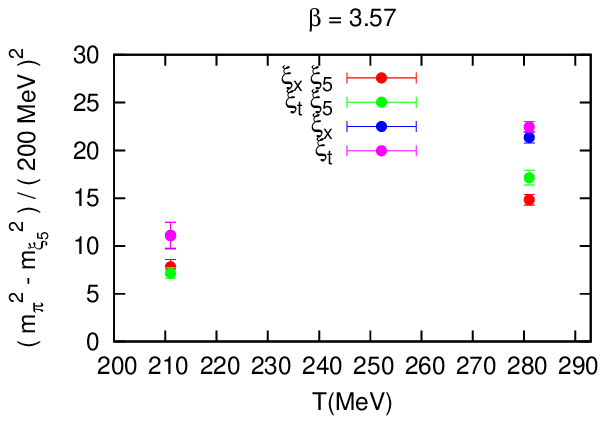}}
\subfigure[]{\label{beta376}\includegraphics[scale=0.9]{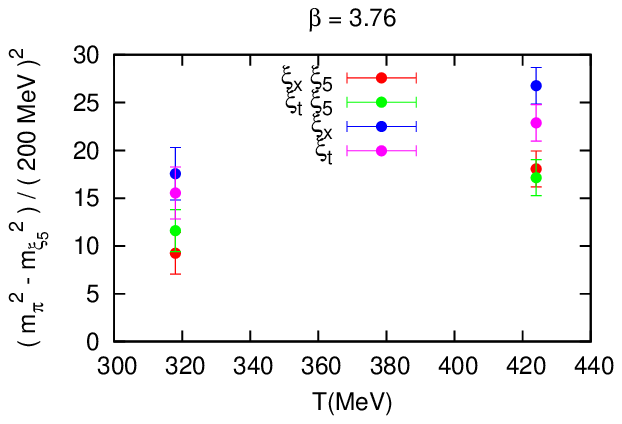}}
\hspace{1cm}
\subfigure[]{\label{beta392}\includegraphics[scale=0.9]{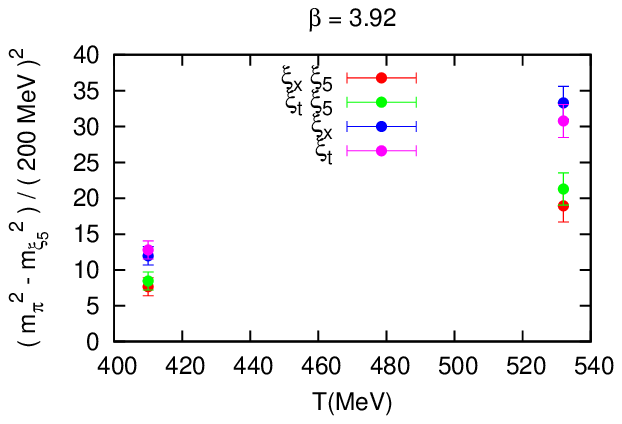}}
\end{center}
\caption{Effect of the temperature on the taste violation at fixed $\beta$ values : (a) $\beta = 3.43$, (b) $\beta = 3.50$, (c) $\beta = 3.57$, (d) $\beta = 3.76$, (e) $\beta = 3.92$ }
\label{fig.betavst}
\end{figure*}

\begin{figure*}[!ht]
\begin{center}
\subfigure[]{\label{t139}\includegraphics[scale=0.9]{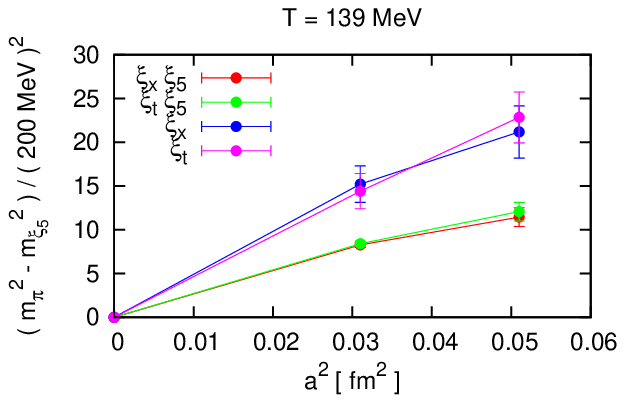}}
\subfigure[]{\label{t175}\includegraphics[scale=0.9]{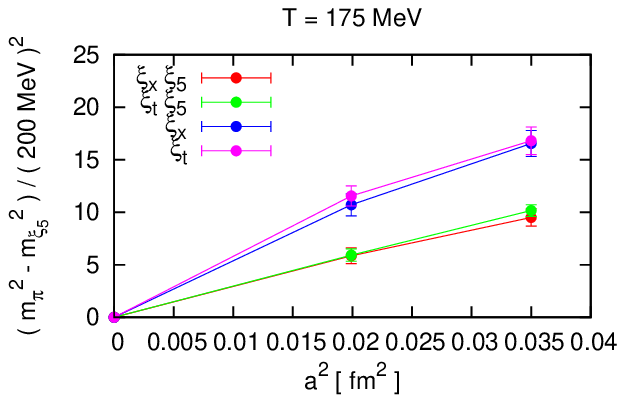}}
\subfigure[]{\label{t226}\includegraphics[scale=0.9]{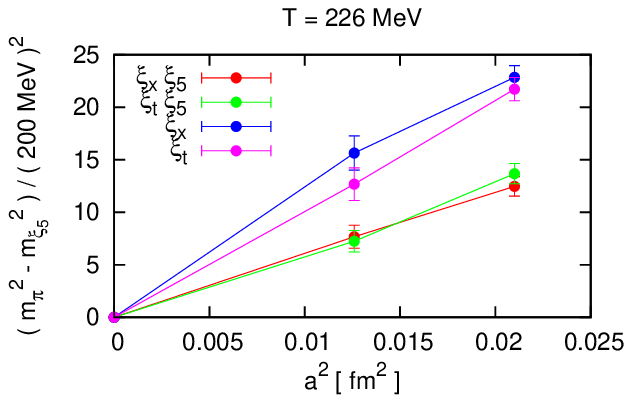}}
\subfigure[]{\label{t318}\includegraphics[scale=0.9]{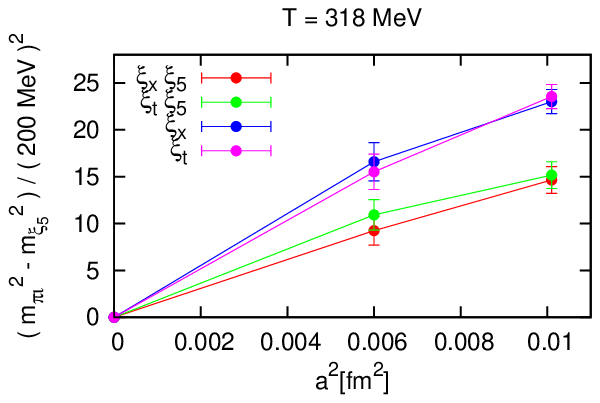}}
\subfigure[]{\label{t361}\includegraphics[scale=0.9]{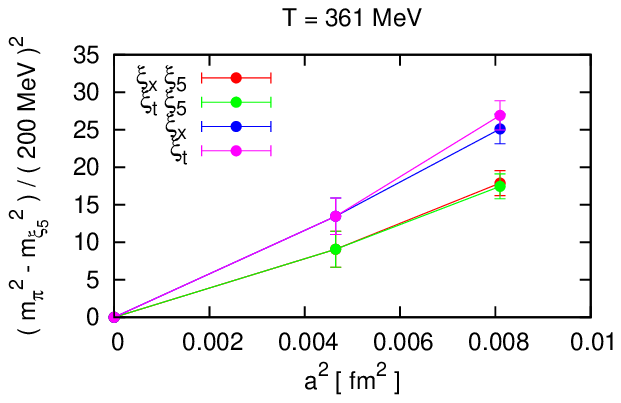}}
\subfigure[]{\label{t549}\includegraphics[scale=0.9]{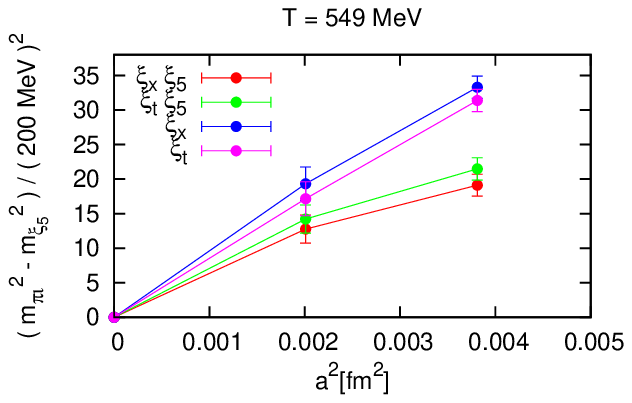}}
\end{center}
\caption{Effect of the lattice spacing on the taste violation at fixed temperature values : (a) T = 139 MeV, (b) T = 175 MeV, (c) T = 226 MeV, (d) T = 361 MeV.}
\label{fig.fff}
\end{figure*}

\begin{figure*}[!ht]
\begin{center}
\subfigure[]{\label{tvsafinalff}\includegraphics[scale=1.05]{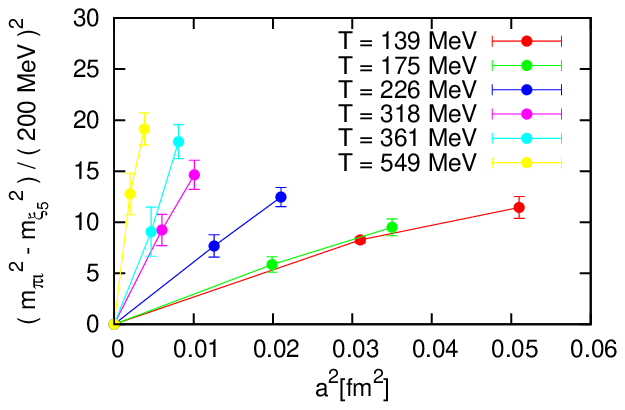}}
\hspace{0.4cm}
\subfigure[]{\label{tvsafinalf}\includegraphics[scale=1.05]{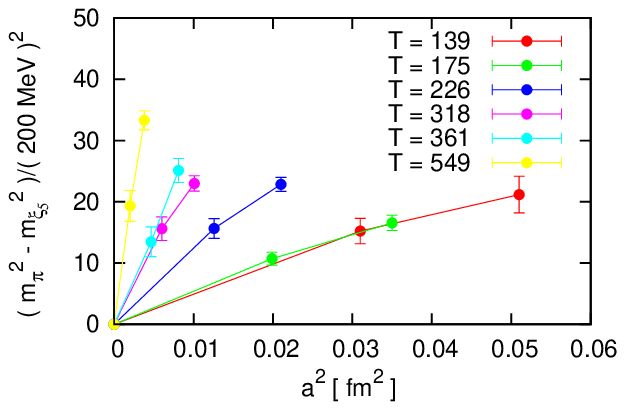}}
\end{center}
\caption{Lattice spacing dependence of the taste splitting for one link ($m_{\gamma_{5}\otimes \xi_x \xi_5}^2 - m_{\gamma_{5}\otimes \xi_5}^2$)(e) and two link meson operators ($m_{\gamma_{5}\otimes \xi_x }^2 - m_{\gamma_{5}\otimes \xi_5}^2$ ) at different temperatures. }
\label{fig.fff2}
\end{figure*}

\begin{figure*}[!ht]
\begin{center}
\subfigure[]{\label{DIFF1}\includegraphics[scale=1.05]{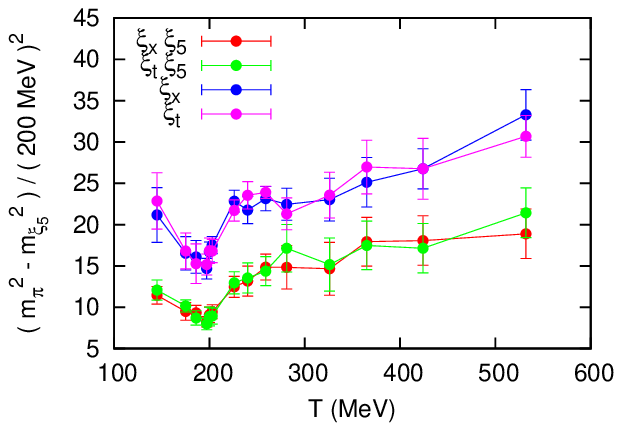}}
\hspace{0.4cm}
\subfigure[]{\label{D2}\includegraphics[scale=1.05]{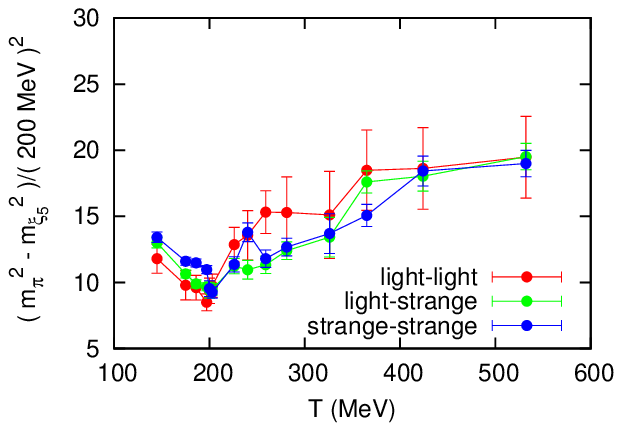}}
\end{center}
\caption{(a) Effect of the temperature on the taste violation in the pion channel
at $N_{\tau}=6$. (b) Flavor dependence of the taste violation for the light-light,
light-strange and strange-strange pseudoscalar channels.
The quantity
$\Delta_{\xi_x} = m_{\gamma_5 \otimes \xi_x}^2 - m_{\gamma_5 \otimes \gamma_5}^2$
is plotted for different flavor content at $N_{\tau}=6$ }
\label{fig.diff}
\end{figure*}

\begin{figure*}[!ht]
\begin{center}
\subfigure[]{\label{fpi}\includegraphics[scale=0.9]{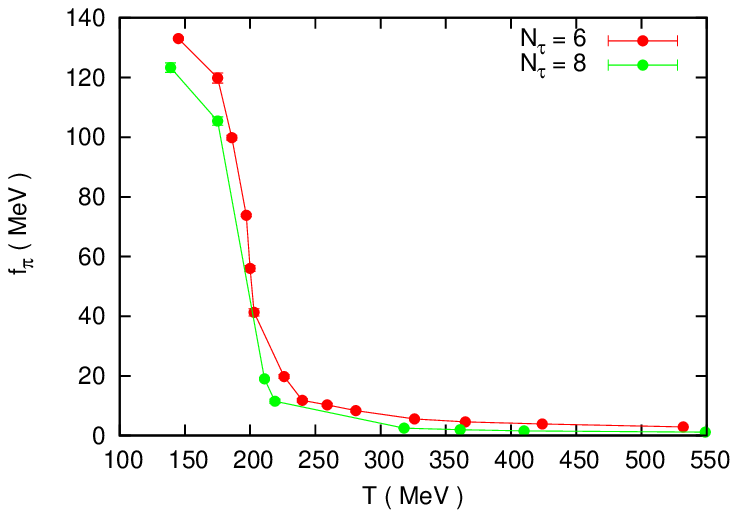}}
\subfigure[]{\label{rho}\includegraphics[scale=0.9]{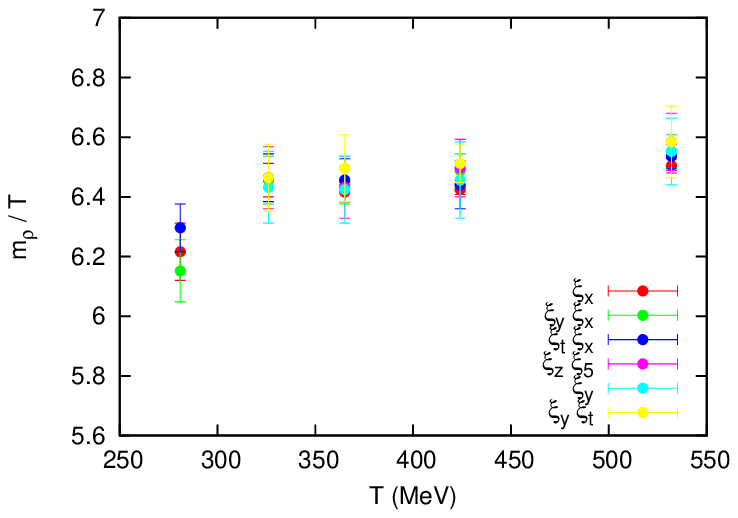}}
\end{center}
\caption{(a) Temperature dependence of the pion decay constant $f_{\pi}$ in MeV calculated at $N_{\tau}$=6. (b) Temperature dependence for different taste state in the vector channel identified by the spin $\times$ taste matrix $\gamma_x \otimes [\xi]$ calculated at $N_{\tau}$=6.}
\label{fig.varie}
\end{figure*}

In order to perform this analysis we have used dynamical $N_f$=2+1 gauge field configurations
generated with the RHMC algorithm \cite{algo} by the RBC-Bielefeld \cite{spatial-string,RBCBi-eos}
and the HotQCD \cite{hotQCD} collaborations using the p4 staggered action. The
configurations were generated along the Line of Constant Physics (LCP) obtained by
tuning the bare quark masses such that at zero temperature the
(Goldstone) pion mass $m_\pi$ is approximately $220$ MeV and the kaon mass $m_K$
equals $500$ MeV. The bare strange quark mass has always been 10 times larger than
the light one, $m_s = 10 m_l$. In table (\ref{tb.runpara}), for both lattices $24^3 \times 6$ and $32^3\times 8$
used in the simulations, the values of the coupling
constant and of the bare light quark mass at which the configurations were
generated are reported.
The number of configurations analyzed was about 500-600 at each temperature, separated
by 10 time units. For more details about the simulations and scale settings we
refer to \cite{RBCBi-eos}.
\newline
In the following we report on our results for the screening masses of the
local and non-local staggered mesons extracted from the lattice calculation
of the spatial correlators in some of the twenty mesonic
channels listed in Table (\ref{tab.Operator}). More in detail
we focus on the pion channels labeled with $n=1,2,7,8,13,14,17,18$
since we are primarily interested in the taste splittings
in this multiplet. However our analysis is not limited to that but we also
inspected some scalar, vector and axial-vector channels to extract
some other useful information from them, for example about the restoration of the chiral
symmetry and the finite temperature splittings of those multiplets.
\newline
Before starting the presentation of the numerical results we recall that a staggered
correlator contains two different mesons with
opposite parity (see eq. \ref{eq.cor-par}). This is valid in general except for the channels $2,7,14,18$
where only the pion contributes. From these channels we are able to extract the screening masses
above and below the transition fitting as usual the correlators with
the exponential functions and with errors determined by jack-knifing the fit. In the
other channels the pions have to be extracted from the oscillating
($n=1,13$) or the non-oscillating part ($n=8,17$) but also the scalars contribute to
the correlators. If one performs the analysis at zero temperature no problem
occurs in the determination of the screening masses of all pion taste components.
Unfortunately the situation is quite different at finite temperature since
in the channels $n=1,8,13,17$ the amplitudes of the pion states 
die out very fast
with rising temperature and as a consequence we are not able to extract
their screening masses. Indeed at temperatures above the transition
the pion contributions to the correlators disappear completely and the channels are dominated by the scalars.
The lack of these screening mass data makes the comparison with the S$\chi$PT
incomplete. Differently from the $T=0$ case in which one can check
an $SO(4)$ restoration of the taste symmetry predicted by the S$\chi$PT,
we are not able to verify if an analogous enhancement
happens also at finite temperature.
\newline
For the axial-vector and vector channels the situation is not so clear, we have seen in all analyzed channels
the presence of both parity states but, likewise to what happens in the
local sector \cite{Cheng:2010fe}, using the point sources we were able to extract only few values of the screening mass for
very high temperatures (above 2 $T_c$) since the correlators are very noisy. \newline
In addition to the investigation done for mesons made out off two degenerate $u$ and $d$ quarks we have
also performed some calculations for meson with flavor content $\bar{u}s$ and $\bar{s}s$
to compare the result with the light sector and extract some additional
checks regarding the S$\chi$PT that predicts the flavor independence of the taste violation.
\newline
\begin{figure*}[!ht]
\begin{center}
\subfigure[]{\label{Pion32r0}\includegraphics[scale=0.9]{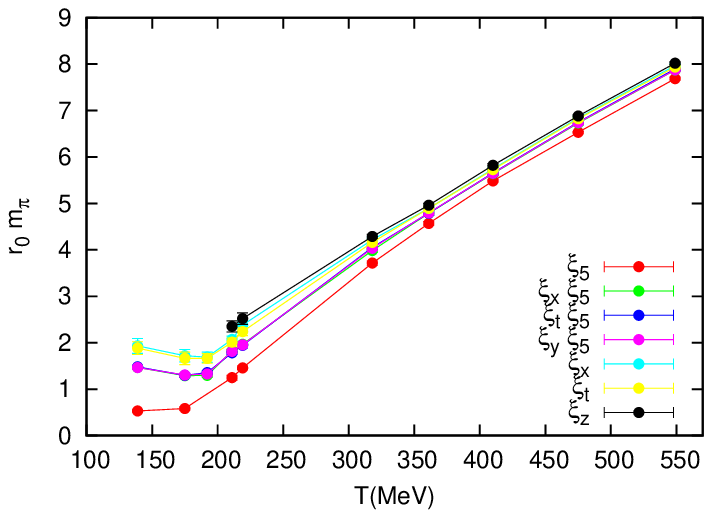}}
\hspace{1cm}
\subfigure[]{\label{Pion32T}\includegraphics[scale=0.9]{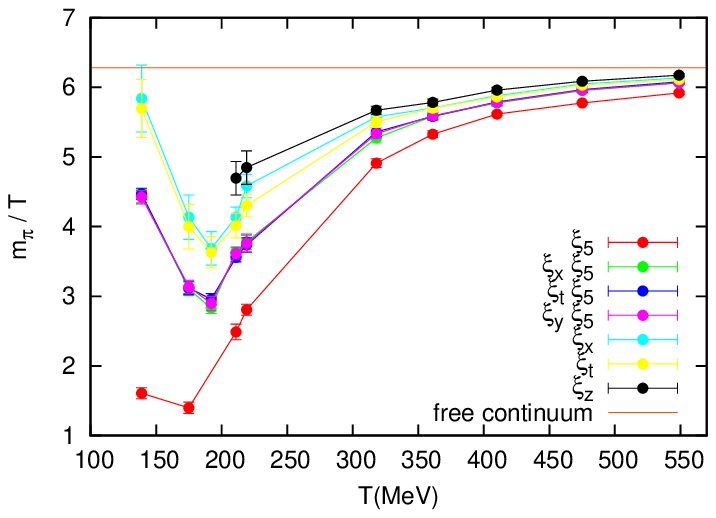}}
\subfigure[]{\label{pion24r0}\includegraphics[scale=0.9]{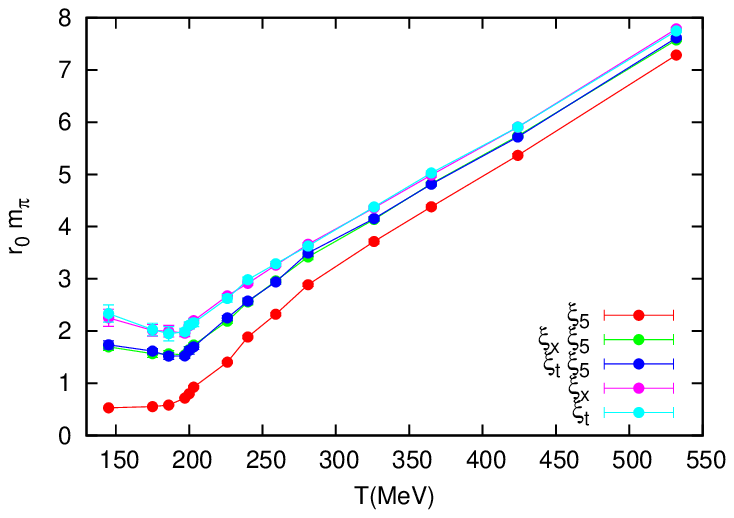}}
\hspace{1cm}
\subfigure[]{\label{pion24T}\includegraphics[scale=0.9]{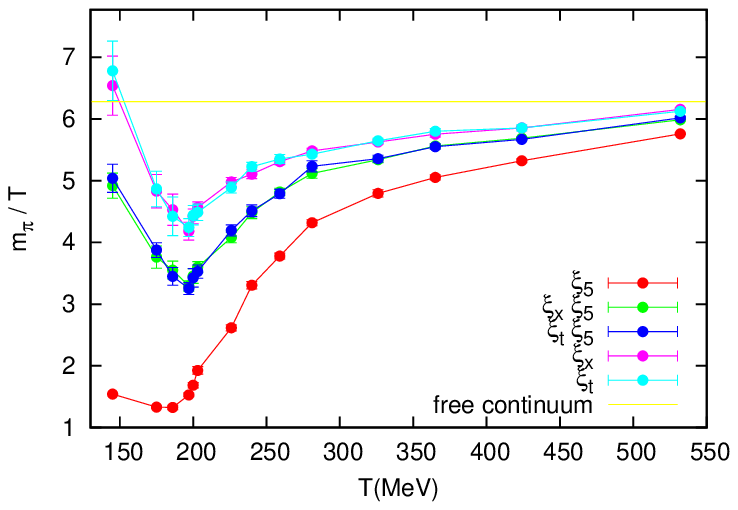}}
\end{center}
\caption{Temperature dependence of the pseudoscalar screening mass ($\gamma_5 \otimes [\xi]$)
in units of $r_0$ and of the temperature at $N_\tau = 8$ $(a,b)$ and $N_\tau = 6$ $(c,d)$ for different components of the multiplet identified by the taste matrix $[\xi]$.}
\label{fig.pion32}
\end{figure*}
Let us start with figure \ref{fig.pion32} where we present the data on the screening mass of the various taste components of the pion
multiplet labeled by their taste matrix $[\xi]$. In particulari, in the pictures \ref{Pion32r0} and \ref{pion24r0} we plot the screening
masses in units of $r_0$ as function of the temperature at $N_{\tau}=8$ and $N_{\tau}=6$ respectively. For both lattices
we can observe that below the transition the Goldstone boson screening mass ($[\xi] = \xi_5$) remains approximatively
constant while for the non local operators a slight decrease of their masses occurs below $T_c$ . Above the transition
there is a rapid linear rise of the values of the masses for all the states with increasing temperature.
From the figures \ref{Pion32T} and \ref{pion24T} where the screening masses in unit of $T$ as function of the temperature
are plotted one can note a strong decrease for the non-local operator masses below the transition and
an increase above that point with an approach from below to the free continuum result given by $2 \pi T$ (indicated by
the continuum line) where the taste symmetry is restored completely.
\newline
\begin{figure*}[!ht]
\begin{center}
\vspace{0.5cm}
\subfigure[]{\label{Scalar32r0}\includegraphics[scale=0.9]{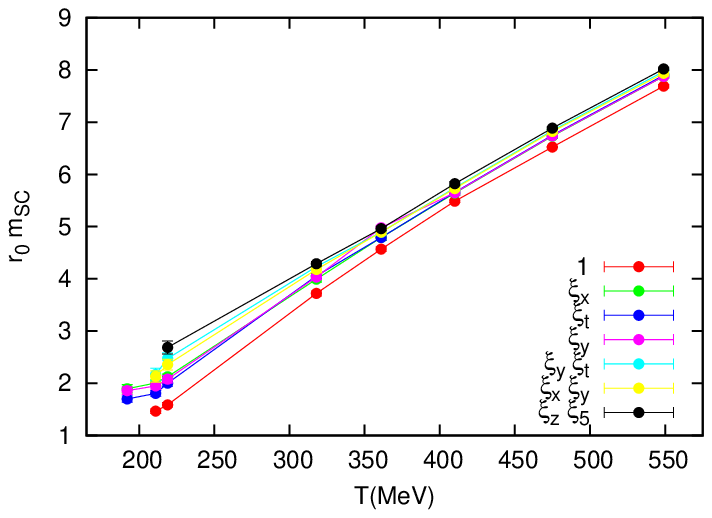}}
\hspace{1cm}
\subfigure[]{\label{Scalar32T}\includegraphics[scale=0.9]{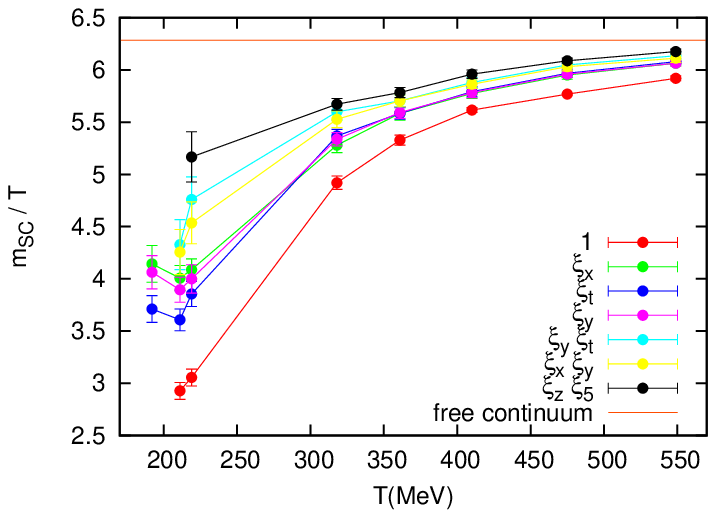}}
\vspace{0.5cm}
\subfigure[]{\label{scalar24r0}\includegraphics[scale=0.9]{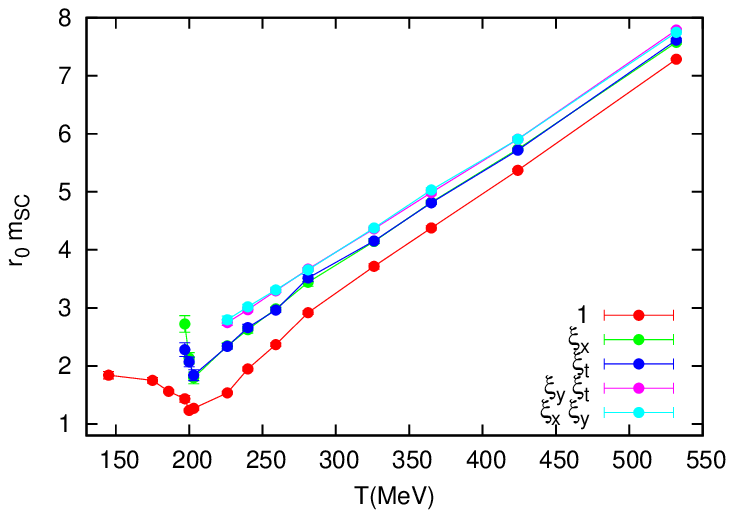}}
\hspace{1cm}
\subfigure[]{\label{scalar24T}\includegraphics[scale=0.9]{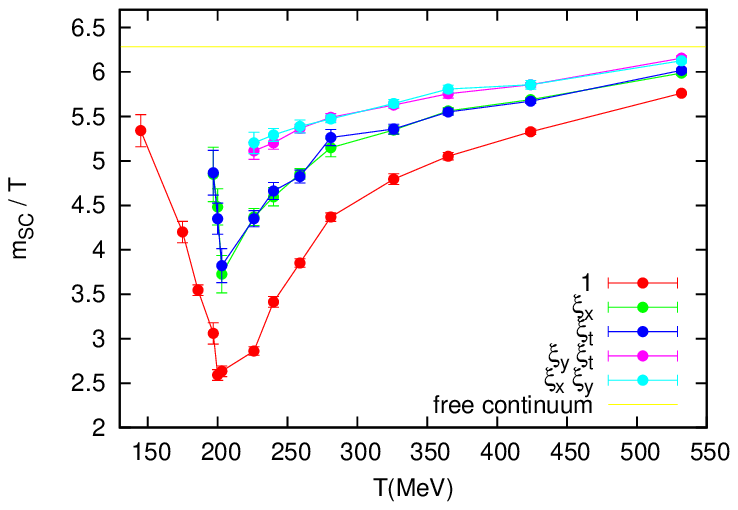}}
\end{center}
\caption{Temperature dependence of the scalar screening mass ($1 \otimes [\xi]$)
in unit of $r_0$ and of the temperature at $N_\tau = 8$ (a,b) and $N_\tau = 6$ (c,d) for different components of the multiplet identified by the taste matrix $[\xi]$.}
\label{fig.scalar32}
\end{figure*}
\noindent
The scalar multiplet is shown in fig. \ref{fig.scalar32} where we plot the temperature dependence of the screening masses of the various taste components normalized by $r_0$ (\ref{Scalar32r0}, \ref{scalar24r0}) and by the temperature (\ref{Scalar32T}, \ref{scalar24T}). Below
the transition it is difficult to obtain quantitative results since the correlators turn out to be very noisy in this region, especially
for the non-local observables. At temperature above $1.2$ $T_c$ all scalar correlators are degenerate with the pseudoscalar ones in corresponding taste channels.
\newline
\begin{figure*}[!ht]
\begin{center}
\subfigure[]{\label{strangelocal}\includegraphics[scale=0.9]{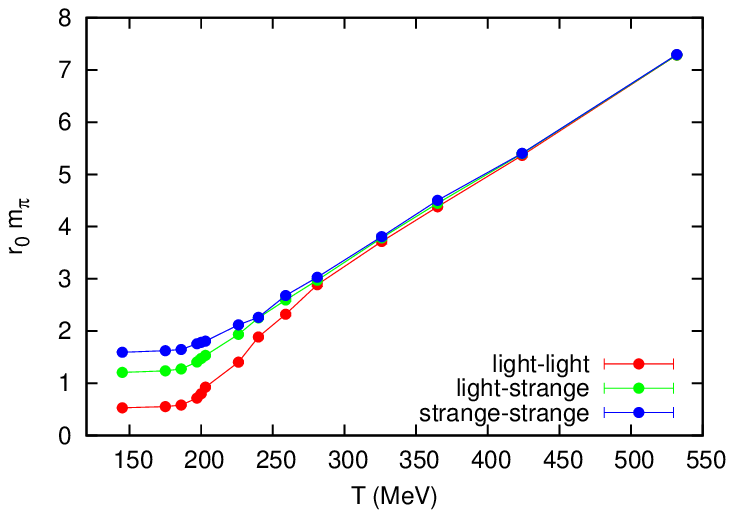}}
\hspace{1cm}
\subfigure[]{\label{strangenlocal1}\includegraphics[scale=0.9]{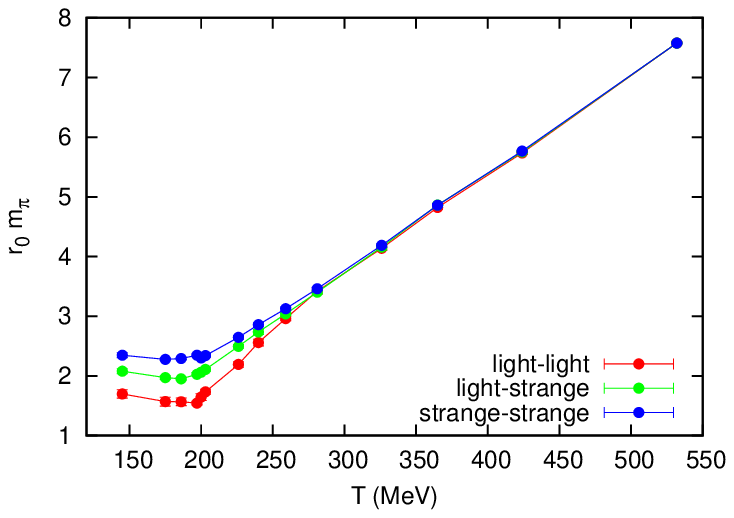}}
\end{center}
\caption{(a) Flavor dependence of the screening mass for the local
pseudoscalar operator ($\gamma_5 \otimes \xi_5$)
in units of $r_0$ at $N_\tau = 6$. (b) Flavor dependence of the screening mass for the 1-link pseudoscalar operator ($\gamma_5 \otimes \xi_x \xi_5$)
in units of $r_0$ at $N_\tau = 6$.}
\label{fig.strange}
\end{figure*}
\noindent
In figure \ref{fig.strange} we show the flavor dependence of the screening mass for the local and the 1-link pseudoscalar channel. More precisely,
we have calculated and plotted the screening masses in the light ($\bar{u}d$), light-strange ($\bar{u}s$) and strange ($\bar{s}s$)
sectors. While above 1.5 $T_c$ for both, local and 1-link operator
no difference between the three sectors can be observed since the quark mass term becomes irrelevant, below $T_c$ Pion, Kaon and $\phi$ split as expected. Furthermore, in the region below the transition we cannot identify any substantial temperature effect on the local operator's masses while for the one-link operators a slight decrease can be observed as the temperature is increased until $T_c$.
Note however that this data has been obtained at fixed $N_\tau$.
\newline
In the following we will discuss the taste splittings which are defined
as the differences of squared pseudoscalar masses of different taste,
see Eq.(\ref{eq.splittings}),
\beq
\Delta_{\xi} = m_{\xi}^2-m_{\xi_5}^2
\eeq
and are traditionally but somewhat arbitrarily normalized to (200 MeV)$^2$.
In order to understand the effect of the temperature on the taste symmetry violation
in the group of figures (\ref{fig.betavst}) we have plotted the taste splittings
as function of $T$
at fixed $\beta$ values. The results indicate that while in the confined phase there is essentially no effect of the temperature on the taste splitting, as one can figure out from figure \ref{beta343}, above the transition a temperature effect can be clearly read from \ref{beta350}, \ref{beta357}, \ref{beta376} and \ref{beta392}. This unexpected behavior above $T_c$ deserves some comments that we postpone to the discussion section. Here we underline the fact and note that this effect seems independent of the lattice spacing since the slope of the increase stays approximatively constant for the configurations at different beta values.
\newline
In the group of figures \ref{fig.fff} we study the lattice spacing dependence of the taste violation at fixed temperature, plotting
the value of the taste splitting as function of $a^2$. For all analyzed temperatures
the taste splittings are compatible with an $a^2$ behavior. 
However as we can easily see from \ref{tvsafinalff} and \ref{tvsafinalf}, while below the transition the slope is more or less constant above $T_c$ its values start to increase with the temperature. This is an indication that in the low temperature regime the predictions of the S$\chi$PT on the taste splitting seem to be valid and independent from $T$; even if above the transition these predictions are no more reliable an $a^2$ behavior is still observed but the temperature starts to affect the taste violation making the quantity $\Delta_{[\xi]}$ temperature-dependent.
\newline
\noindent
In figure \ref{DIFF1} we have plotted the taste splitting for different taste states in units of 200 MeV as function of the
temperature. The dependence of the taste violation cannot be extracted trivially looking at this pictures since in the plot we are comparing
different finite temperature configurations with the same $N_{\tau}$ but thus with
different values of the lattice spacing. For
example comparing this figure with \ref{tvsafinalff} and \ref{tvsafinalf} one can note that the strong decreasing of the taste
splitting observed below $T_c$ in (\ref{DIFF1}) and (\ref{D2}) is just a pure discretization
effect. Above $T_c$ the behavior of the splittings changes drastically, first starting
to rise rapidly and at 1.5 $T_c$ there is a flattening of the slope but again this is
a combined lattice spacing and temperature effect.
\newline
The flavor dependence on the taste violation is analyzed in the figure \ref{D2} where
the quantity
$\Delta_{\xi_x} = m_{\gamma_5 \otimes \xi_x}^2 - m_{\gamma_5 \otimes \gamma_5}^2$
for the $\bar{u}d$, $\bar{u}s$ and  $\bar{s}s$ channel is plotted. In the range of the
temperature analyzed the screening masses in the different sectors are in agreement
within the error bars. But while below the transition where the quark mass
is relevant this fact is a not-trivial prediction of the staggered chiral perturbation
theory, above 1.5 $T_c$ the plot tells us simply that
as noted in fig. \ref{strangelocal} and \ref{strangenlocal1} the effect of the quark
mass is negligible.
\newline
In order to study the restoration of the chiral symmetry and the breakdown of
the chiral perturbation theory it's interesting to study the temperature dependence
of the pion decay constant $f_{\pi}$ that is defined on the
lattice \cite{Kilcup:1986dg} as

\begin{equation}f_{\pi} = \frac{1}{4}\, \frac{m_{l}\, \,   \sinh{(m_{\pi})\, }^{1/2}}{ \sinh{(m_{\pi}/ 2  ) }^{2}} \sqrt{A_{PS}}\end{equation}
\noindent
where $m_{\pi}$ and $A_{PS}$\footnote{The amplitude $A_{PS}$ is related to $A_{NO(O)}$ (equation \ref{eq.cor-par}) trough the relation $A_{NO(O)} = 2 A_{PS} e^{-m_{\pi} N_s /2} $} are respectively the mass and the amplitude extracted from the Goldstone pion correlator. This
relation is nothing else than the staggered lattice version \cite{Kilcup:1986dg} of the continuum definition through
the coupling of the axial current to the pion

\begin{equation}\sqrt{2} f_{\pi} m_{\pi} = - \langle 0| \bar{u} \gamma_4 \gamma_5 d | \pi^+ \rangle.\end{equation}
\noindent
From figure \ref{fpi} one can essentially note the sharp decrease of the value of $f_{\pi}$ in the proximity of the
transition and a near vanishing after that point. This signals clearly that the chiral symmetry is restored and the chiral perturbation
theory is no more applicable. In the chiral symmetry broken phase even if only few data are available we can observe a temperature
dependence of $f_{\pi}(T)$. Another thing to point out is the difference between
the $N_{\tau}=8$ and the $N_{\tau}=6$ results. This is
probably due to the fact that in the first case the
discretization errors are reduced compared to the second
lattice and indeed the values are closer to the
continuum value of $f_{\pi}(0) \simeq 93 $ MeV. It would be
interesting to
get some more data in the confined phase and to perform an extrapolation to the physical mass point and to the continuum similar to that done at zero temperature in \cite{Bazavov:2009bb}.
\newline
In the last picture we plot the temperature dependence of the screening masses for some taste states in the vector channel. Unfortunately
since at low temperatures these correlators are very noisy it is in general difficult to
extract the vector screening masses for $T < 2 T_c$. Looking at figure \ref{rho} we can
essentially note two facts: the first is that at temperatures between 2 and 3 $T_c$ the
values of the screening masses in the vector channel are more or less constant and have
already reached a value of $2 \pi T$ which is the infinite temperature
value in the thermodynamic limit. The second point is that taste violations do not seem
to be present in these channels since the difference between the various tastes are
within the error bars. We have also looked at other vector and axial-vector channels finding
 analogous results.\newline

\section{Summary}               \label{se.summary}

In this paper we have investigated the violation of taste symmetry at finite temperature
in $2+1$ flavor QCD  utilizing the gauge configurations generated by the RBC-Bielefeld
\cite{spatial-string,RBCBi-eos} and the HotQCD \cite{hotQCD} collaborations using the
improved p4 fermion action. Two different lattice sizes, $24^3 \times 6$ and $32^3 \times 8$,
have been used to understand the role
of the temperature in the taste breaking phenomena. We have calculated the screening masses by fitting
the spatial meson correlators and even if we have been primarily concentrated
on the pion channels, some other mesonic states built from
strange quarks and some scalar, axial-vector and vector operators have been considered.\newline
The analysis of the taste symmetry breaking is important since this violation
is related to the cut-off dependence of the physical quantities.
At temperatures below the transition our aim was to understand whether the predictions
of the staggered chiral perturbation theory on the taste violation are still valid, with possibly
temperature dependent effective low energy constants, or whether eventually some new terms in the chiral lagrangian would have to be considered. As pointed out in the previous section
for $T <T_c$ we have found indications that the taste splitting in the pion channel has an
$a^2$ behavior as predicted by S$\chi$PT and no effects on the temperature
can be observed in this region. 
It could be interesting to compare the results with zero temperature
taste splittings with the same action.\newline
The apparent temperature independence of the taste splittings in combination
with a noticeably temperature dependent pion decay
constant (fig. \ref{fpi}) 
implies within staggered chiral perturbation
theory that the coefficients $C_i$ of the lagrangian (\ref{chila})
acquire a temperature dependence.
However, as not all taste states in the pion multiplet are accessible at
high temperature due to their dying out when the transition region
is approached, the system of equations for the $C_i$ cannot be solved
and it is not possible to determine the
temperature behavior of these LECs.\newline
Above the transition the situation changes drastically. As also indicated
by the vanishing of the pion decay constant
$f_{\pi}$, chiral symmetry is restored and
the predictions of chiral perturbation theory are no longer valid.
Staying nevertheless with the chirally inspired definition of
taste splittings, eq. (\ref{eq.splittings}), this quantity measures
taste splittings
that behave like $a^2$ - as is expected - but with a slope which increases
with temperature, \emph{i.e.} the splittings appear to be temperature dependent.
However, because chiral symmetry is restored in this region the pion
ceases to be a Goldstone particle and acquires a non-vanishing mass.
For sufficiently high temperatures the screening masses will become
proportional to the temperature. As a consequence, the splittings as defined in
(\ref{eq.splittings}) should then behave according to
\begin{equation}\Delta_{[\xi]} \sim  m_{\xi_5} (  m_{[\xi]} - m_{\xi_5} )
\sim T (  m_{[\xi]} - m_{\xi_5} ) \gtrsim a^2 T,\end{equation}
\noindent
\emph{i.e.} rise with temperature.
The temperature interval investigated,
$T_c < T < 2\, T_c$, represents the transition region where the pion
changes from Goldstone nature to approaching free theory behavior such
that the temperature dependence is more complicated and a linear rise not yet
clearly identifiable. It would be interesting to better understand 
the behavior in the chirally restored
phase and to compare for instance with perturbative predictions which could guide
the numerical analysis and should be reliable at high temperature.\newline
Another point that we have analyzed carefully is the finite temperature
splitting of the $T=0$ multiplets \cite{Gupta:1999hp}.
According to group theory \emph{e.g.} the triplets of the $T=0$ GTS
split into a two plus
a one dimensional irrep of the finite temperature GTS.
As was already noted in the literature \cite{Cheng:2010fe} for some
local axial-vector
and vector channels a splitting between "longitudinal" and "transverse"
screening masses has been observed. 
Indeed, also in the other, non-local
triplet channels ($n=5,6,11,12,19,20$) these splittings can be seen even
if the analysis is performed with point source
operators where only few data in the high $T$ regime are available.
The situation may improve with the use of
wall sources succeeding to analyze configurations at lower temperatures.
Quite surprisingly, however, we found that pion triplets $n=7, 8, 13, 14$
do not split in this way. Indeed the difference between
the "transversal" and the "longitudinal" screening masses is compatible
within the error bars.
This is a clear indication that at finite $T$ at least in the pion channel,
contrary to group theory arguments,
an enhancement of the
$SO(2) \times Z_2$ symmetry
occurs and a restoration of an $SO(3)$ symmetry
can be observed in these channels.
As already stressed previously we don't have access
to all the taste masses, so that we cannot check for a complete $SO(4)$
restoration like in the zero temperature case.\newline
Finally we have evaluated the flavor effect on the taste splitting.
For temperatures above 1.5 $T_c$ the splittings are flavor independent
for the simple reason that in this region the effect of the quark
mass is negligible. For temperatures below the transition the fact that
no flavor effect can be observed is non-trivial
and supports the applicability of staggered chiral perturbation theory,
which predicts that the pion splittings are independent of the flavor of
the quark constituents, in this temperature regime.

\section*{Acknowledgments}

This work is supported by the Research Executive Agency (REA)
of the European Union under Grant Agreement PITNGA-
2009-238353 (ITN STRONGnet). F.P. would like to thank the
hospitality of the G. Galilei Institute for Theoretical
Physics, Florence. The numerical
computations have been carried out on the apeNEXT at Bielefeld University.

\section*{Appendix}

In the tables \ref{33} and \ref{tb.nM83} we have summarized the pion screening
masses for different taste states. The screening masses are given in
lattice units and the data can be easily converted to either temperature or vacuum
($r_0$) units by means of the included zero temperature results for $r_0/a$ from
\cite{RBCBi-eos}.

\begin{table*}
\begin{center}
\begin{tabular}{|lr|cccccc|}
\hline
$\beta$ & $r_0/a$ & $a M_{PS}$  & $a M_{\pi}$  & $a M_{\pi}$  & $a M_{\pi}$ & $a M_{\pi}$ & $a M_{\pi}$\\
     &   & [$\gamma_5\otimes \xi_5$] &  [$\gamma_5\otimes \xi_x \xi_5$] & [$\gamma_5\otimes \xi_t \xi_5$] & [$\gamma_5\otimes \xi_x $] &
     [$\gamma_5\otimes \xi_t$] & [$\gamma_5\otimes 1 $] \\
\hline
\hline
\multicolumn{8}{|c|}{} \\
\multicolumn{8}{|c|}{$32^3 \times 8$} \\
\hline
3.430 &  2.647 & 0.201(01) & 0.555(12) & 0.559(10) & 0.730(60) & 0.712(52) & -        \\
3.500 &  3.328 & 0.175(01) & 0.382(12) & 0.390(12) & 0.517(40) & 0.500(40) & -        \\
3.530 &  3.654 &    -      & 0.354(10) & 0.370(12) & 0.461(30) & 0.454(28) & - \\
3.570 &  4.009 & 0.311(14) & 0.455(09) & 0.445(08) & 0.517(18) & 0.503(22) & 0.587(30) \\
3.585 &  4.160 & 0.351(10) & 0.473(14) & 0.467(13) & 0.573(21) & 0.538(23) & 0.606(30) \\
3.760 &  6.050 & 0.614(08) & 0.659(07) & 0.670(08) & 0.697(10) & 0.688(10) & 0.709(07) \\
3.820 &  6.864 & 0.666(06) & 0.698(07) & 0.698(06) & 0.713(07) & 0.713(06) & 0.723(06) \\
3.920 &  7.814 & 0.702(03) & 0.722(06) & 0.724(06) & 0.735(05) & 0.733(04) & 0.745(05) \\
4.000 &  9.048 & 0.721(02) & 0.744(03) & 0.746(02) & 0.756(03) & 0.754(03) & 0.761(03)\\
4.080 & 10.390 & 0.740(03) & 0.758(03) & 0.759(03) & 0.767(03) & 0.764(02) & 0.772(03) \\
\hline
\end{tabular}
\end{center}
\caption{Screening masses for some taste states in the pion multiplet from $N_\tau = 8$ lattices.
}
\label{33}
\end{table*}

\begin{table*}
\begin{center}
\begin{tabular}{|lr|ccccc|}
\hline
$\beta$ & $r_0/a$ & $a M_{PS}$  & $a M_{\pi}$  & $a M_{\pi}$  & $a M_{\pi}$ & $a M_{\pi}$ \\
     &   & [$\gamma_5\otimes \xi_5$] &  [$\gamma_5\otimes \xi_x \xi_5$] & [$\gamma_5\otimes \xi_t \xi_5$] & [$\gamma_5\otimes \xi_x $] &
     [$\gamma_5\otimes \xi_t$] \\
\hline
\hline
\multicolumn{7}{|c|}{} \\
\multicolumn{7}{|c|}{$24^3 \times 6$} \\
\hline
3.351 &  2.069 & 0.257(01) & 0.820(34) & 0.840(38) & 1.090(80) & 1.130(80)   \\
3.410 &  2.503 & 0.222(03) & 0.627(30) & 0.646(20) & 0.805(45) & 0.811(48)   \\
3.430 &  2.647 & 0.221(06) & 0.592(24) & 0.575(24) & 0.755(42) & 0.737(52)   \\ 
3.445 &  2.813 & 0.255(07) & 0.549(14) & 0.542(16) & 0.697(24) & 0.707(24)   \\
3.455 &  2.856 & 0.281(07) & 0.576(20) & 0.571(24) & 0.737(20) & 0.740(26)   \\
3.460 &  2.890 & 0.321(10) & 0.600(15) & 0.588(18) & 0.761(15) & 0.748(22)   \\
3.490 &  3.223 & 0.436(08) & 0.680(14) & 0.688(15) & 0.830(12) & 0.815(10)   \\
3.510 &  3.426 & 0.551(10) & 0.747(16) & 0.752(16) & 0.851(12) & 0.871(12)   \\
3.540 &  3.687 & 0.630(08) & 0.803(10) & 0.798(12) & 0.885(08) & 0.892(12)  \\
3.570 &  4.009 & 0.720(08) & 0.853(13) & 0.872(15) & 0.914(10) & 0.905(07)  \\
3.630 &  4.651 & 0.799(10) & 0.890(08) & 0.893(09) & 0.938(07) & 0.941(07)  \\
3.690 &  5.201 & 0.842(07) & 0.927(06) & 0.925(05) & 0.959(08) & 0.967(07)  \\
3.760 &  6.050 & 0.887(04) & 0.948(03) & 0.945(03) & 0.976(04) & 0.976(03)  \\
3.920 &  7.590 & 0.960(02) & 0.998(03) & 1.003(04) & 1.026(04) & 1.021(04)  \\
\hline
\end{tabular}
\end{center}
\caption{Screening masses for some taste states in the pion multiplet from $N_\tau = 6$ lattices.
}
\label{tb.nM83}
\end{table*}


\end{document}